\documentclass[twocolumn]{aastex63}

\received{XXXX}
\revised{XXXX}
\accepted{XXXX YY, 2022}
\submitjournal{AJ}
\shorttitle{A Silent Revolution in Fundamental Astrophysics}
\shortauthors{Eker et al.}
\usepackage{xcolor}
\begin{document}

\title{A Silent Revolution in Fundamental Astrophysics}

\correspondingauthor{Zeki Eker}
\email{eker@akdeniz.edu.tr}

\author[0000-0003-1883-6255]{Zeki Eker}
\affiliation{Akdeniz University, Faculty of Sciences, Department of Space Sciences and 
Technologies, 07058, Antalya, Turkey}

\author[0000-0002-5141-7645]{Faruk Soydugan}
\affiliation{Department of Physics, Faculty of Arts and Sciences, \c{C}anakkale Onsekiz 
Mart University, 17100 \c{C}anakkale, Turkey}
\affiliation{Astrophysics Research Center and Ulup{\i}nar Observatory, \c{C}anakkale 
Onsekiz Mart University, 17100, \c{C}anakkale, Turkey}

\author[0000-0002-3125-9010]{Volkan Bak{\i}\c{s}}
\affiliation{Akdeniz University, Faculty of Sciences, Department of Space Sciences and 
Technologies, 07058, Antalya, Turkey}

\author[0000-0003-3510-1509]{Sel\c{c}uk Bilir}
\affiliation{Istanbul University, Faculty of Science, Department of Astronomy and Space Sciences, 34119, Beyaz\i t, Istanbul, Turkey}

\author[0000-0003-3716-858X]{Ian Steer}
\affiliation{176 The Esplanade, Ste. 705, Toronto, Canada M5A 4H2}

\begin{abstract}
Arbitrariness in the zero point of bolometric corrections is a nearly century old paradigm leading to two more paradigms. ``Bolometric corrections must always be negative,'' and ``bolometric magnitude of a star ought to be brighter than its $V$ magnitude''. Both were considered valid before IAU 2015 General Assembly Resolution B2, a revolutionary document that supersedes all three aforementioned paradigms. The purpose of this article is to initiate a new insight and a new understanding of the fundamental astrophysics and present new capabilities to obtain standard and more accurate stellar luminosities and gain more from accurate observations in the era after {\it Gaia}. The accuracy gained will aid in advancing stellar structure and evolution theories, and also Galactic and extragalactic research, observational cosmology and dark matter and dark energy searches.  
 
\end{abstract}

\section{Introduction}
\label{sec:introduction}
According to \citet{Kuhn62}, one of the preeminent philosophers of science in the 20th century, scientists work under paradigms between scientific revolutions. This is the period when scientists tend to ignore anomalies opposing the dominant paradigm. It is only when odds are too much to hide that a crisis and then a revolution occur before the next normal science period begins and a new paradigm guides scientists. The modern history of science attests to two major revolutions separating three such periods. First, the Copernican revolution advanced science from a geocentric to a heliocentric view, although both models could be considered static. Second, Einstein’s theories of special and general relativity advanced science from a static to a dynamical (expanding) model. The discovery of Hubble’s law \citep{Hubble29} led to a major revolution in observational astrophysics and cosmology and the new world picture of a universe undergoing expansion. One must also keep in mind however, that every branch of science has its minor as well as major revolutions. They can include less noticed and even silent revolutions.

The discover that the expansion of the universe is apparently accelerating, based on observations of Type Ia supernovae, was a surprise that led to a partial revolution in cosmology \citep{Perlmutter98, Riess98, Schmidt98}. Furthermore, a crisis in cosmology known as Hubble tension has developed. The Hubble constant ($H_0$) or universal expansion rate determined from the cosmic microwave background (CMB) of $67.4\pm0.5$ km s$^{-1}$ Mpc$^{-1}$ \citep{Planck20}, does not agree with the rate determined from Cepheid-calibrated Type Ia supernova, at $73.04\pm0.04$ km s$^{-1}$ Mpc$^{-1}$ \citep{Riess22}. The $4\sigma-5\sigma$ disparity appears potentially intractable because it continues to increase in severity as the sensitivity of measurements gets better \citep{Melia22}. \citet{Melia22} also thinks perhaps the observations may not be fully consistent with an accelerating universe. Even if it were so, the incompatibility of the Hubble rates noted cannot be explained by simply replacing the currently accepted cosmological model. 

For us the problem appears more fundamental than that. That is, there may not be an immediate solution but rather a marathon of revising current extragalactic distances and distance indicators and aiming to improve various methods of measuring $H_0$ in order to be consistent with the rate according to CMB observations. One of the intolerable negative effects of the paradigm of arbitrariness attributed to the zero point of the bolometric correction (BC) scale is the extra uncertainty in the predicted luminosity ($L$) caused by different zero points preferred by different authors in the literature for more than 80 years. It was first discussed by \citet{Torres10} that this uncertainty could be about 10\% or larger. It is named the zero-point error caused by arbitrariness by \citet{Eker21a, Eker21b}, who not only confirmed \citet{Torres10} but also claimed that it is possible to avoid it if a standard BC is used when computing standard $L$ for a star. Being equivalent to a 5\% error in parallax, a systematic 10\% zero-point error in a predicted $L$ in addition to its uncertainly caused by random observational errors associated with absolute magnitude $M_{\zeta}$ and ${\rm BC}_{\zeta}$ of a star, where $\zeta$ could be any photometric band, is intolerable nowadays, especially after {\it Gaia}. 

We suspect it is not a simple coincidence that 10\% systematic error in predicted $L$ is on par with the discrepancy involved in the Hubble tension \citep{Planck20, Riess22}. Improving the accuracy of stellar luminosities is the starting point to improving galactic and extragalactic luminosities, luminosity functions, and distances because galaxies are mainly made of stars. Accurate determination of luminous mass in galaxies and galaxy clusters would enable us to determine dark matter more accurately. On the other hand, improvements in extragalactic luminosities enable improved extragalactic distance estimates. Thus, cosmological models and the value of the Hubble constant cannot be independent of current improvement in fundamental astrophysics. One step has already been taken to avoid the legacy problems caused by the arbitrariness of the BC scale by defining a truly standard BC and standard $L$ \citep{Eker21a, Eker21b}. It has also been discussed and shown that standard stellar luminosities with accuracies at 1\% are now possible \citep{Eker21b, Bakis22}.

IAU 2015 General Assembly Resolution B2 is a revolutionary document in part because it has the potential to start a new revolution in astrophysics and to aid researchers in solving the current crisis in cosmology. First, it discards the paradigm of arbitrariness in the zero point of bolometric magnitudes. It furthermore discards two following paradigms. ``Bolometric corrections must always be negative,'' and ``bolometric magnitude of a star ought to be brighter than its $V$ magnitude''.

We see IAU 2015 General Assembly Resolution B2 as a silent and so far unnoticed revolution. We believe its full potential is far from being realized. One may note that there are numerous articles in respected journals that continue to ignore the resolution, which we do not reference directly so as not to offend. Evidently, the three paradigms involved are so basic that many authors prefer to ignore the effects as trivial anomalies, typical of normal science. In this study, we show how and why adoption of the precepts, advantages, and advancements in the resolution will improve fundamental astrophysics, and as a result advance modern cosmology and extragalactic as well as Galactic and stellar research. 

\section{Breaking Paradigms of Fundamental Astrophysics}

The first paradigm broken was the arbitrariness attributed to the zero-point constant of the BC scale. IAU 2015 General Assembly Resolution B2\footnote{${\rm https://www.iau.org/static/resolutions/IAU2015\_English.pdf}$} (hereafter IAU 2015 GAR B2) superseded this paradigm by issuing: 

\begin{equation}
M_{\rm Bol}= -2.5\times\log(L/L_{0})=-2.5\times\log L + 71.197~425~...
\end{equation} 
where $L_{0}=3.0128\times 10^{28}$ W is the radiative luminosity of a star with absolute bolometric magnitude $M_{\rm Bol}=0$ mag. It corresponds to the value of the zero-point constant $C_{\rm Bol}=71.197~425~...$ mag if the star's $L$ is in SI units. Then, the following relation gives absolute bolometric magnitude $M_{\rm Bol}$ of a source of luminosity $L$ expressed in SI unit W.  

\begin{equation}
M_{\rm Bol}= -2.5\times\log L + C_{\rm Bol}.
\end{equation} 

If $L$ is replaced by $L_{\rm V}$, which is $V$ filtered luminosity of the same star, it can be written for its visual absolute magnitudes as:
\begin{equation}
M_{\rm V}= -2.5\times\log L_{\rm V} + C_{\rm V}.
\end{equation} 

Since $L_{\rm V}$ of a star is less than its $L$, the zero-point constant for the $V$ filter ($C_{\rm V}$) must be smaller than $C_{\rm Bol}$. Subtracting equation (3) from (2),  

\begin{equation}
{\rm BC}_{\rm V}= M_{\rm Bol}-M_{\rm V} = 2.5\times\log (L_{\rm V}/L) + C_{\rm Bol} - C_{\rm V}.
\end{equation} 
and identifying $C_2 = C_{\rm Bol}- C_{\rm V}$ as the zero-point constant of the $V$ band bolometric corrections ${\rm BC}_{\rm V}$, it is obvious that $C_2$ is a positive number because both $C_{\rm Bol}$ and $C_{\rm V}$ are positive and $C_{\rm Bol} > C$.

Equation (4) indicates that the zero point of ${\rm BC}_{\rm V}$ is not arbitrary because both $C_{\rm Bol}$ and $C_{\rm V}$ are well-defined constants and the zero-point constant is equal to their difference. Therefore, the first superseded paradigm is ``the zero-point constant of the bolometric correction scale is arbitrary''. 

Like a chain reaction, the other two paradigms ``bolometric corrections must always be negative'' and ``bolometric magnitude of a star ought to be brighter than its visual magnitude'' are also superseded. Since $L_{\rm V}/L$ is a number between zero and one ($0<L_{\rm V}/L< 1$) indicating $\log(L_{\rm V}/L$) is a negative number while $C_2$ is a positive number, then ${\rm BC}_{\rm V}$ is positive if the absolute value of the logarithmic term is smaller than $C_2$ or else it is negative. Note: according to (4), ${\rm BC}_{\rm V} = C_2$ and ${\rm BC}_{\rm V} > C_2$ are not allowed. Positive BC is also possible by the following equation derived from (4)
\begin{equation}
L_{\rm V}= L\times 10^{(BC_{\rm V}-C_2)/2.5}.
\end{equation} 

Since it is unphysical to have $L_{\rm V}=L$ or $L_{\rm V}>L$, equation (5) indicates all ${\rm BC}_{\rm V}$ values less than $C_2$ are valid. In other words, not only ${\rm BC}_{\rm V}<0$, but also $0<{\rm BC}_{\rm V}<C_2$ are valid to produce $L_{\rm V}<L$. Therefore, BCs cannot be limited only to negative numbers. There could be stars with positive BC \citep[see][]{Eker20, Eker21a, Eker21b} satisfying the condition $L_{\rm V}<L$. Main-sequence stars having effective temperatures between 5859 K and 8226 K with positive ${\rm BC}_{\rm V}$ have indeed been found \citep{Eker20}.  

Even a single occurrence of a positive BC (regardless of its value) breaks the paradigm that ``bolometric magnitude of a star ought to be brighter than its $V$ magnitude'' because a positive BC indicates $M_{\rm Bol}>M_{\rm V}$, which is an obvious case sufficient to break this paradigm if the word ``brighter'' mean smaller number. This paradigm is also incorrect factually and linguistically since eye comparisons of magnitudes are meaningful only if they are done in the same wavelength range within the visible part of the electromagnetic spectrum; otherwise, the comparison is meaningless due to different eye sensitivity at different wavelengths. One final comment: BC and colors, e.g. $U-B$, $B-V$, $V-R$, $V-I$ ... etc, are not defined to compare the brightness of a star at different bands, they are rather defined as physical entities to reveal information about effective temperature and/or Spectral Energy Distribution (SED) of stars. 

\section{Superseding the Paradigms in Apparent Magnitudes}

IAU 2015 GAR B2 did more than fix the zero point of absolute bolometric magnitudes. It also fixed the zero point of apparent bolometric magnitudes, because fixing the zero point of absolute bolometric magnitudes (equation 2) automatically fixes the zero point of apparent bolometric magnitudes. It is also written in the same document: “the zero point of the apparent bolometric magnitude scale by specifying that $m_{\rm Bol}=0$ mag corresponds to an irradiance or heat flux density of $f_0=2.518021002~...~\times 10^{-8}$ W m$^{-2}$ and hence the apparent bolometric magnitude $m_{\rm Bol}$ for an irradiance $f$ (in W m$^{-2}$) is

\begin{equation}
m_{\rm Bol}= -2.5\times\log (f/f_0) = -2.5\times\log f - 18.997~351~...
\end{equation} 
The irradiance $f_0$ corresponds to that from an isotropically emitting radiation source with absolute bolometric magnitude $M_{\rm Bol}=0$ mag (luminosity $L_{0}$) at a standard distance of 10 parsecs'' (IAU 2015 GAR B2). Equation (6) also reveals that the zero-point constant of the apparent bolometric magnitude scale is no longer arbitrary as well. Thus, its value is: $c_{\rm Bol}({\rm irr}) = -18.997~351~...$ mag, then  

\begin{equation}
m_{\rm Bol}= -2.5\times\log (f_{\rm Bol}) + c_{\rm Bol}(\rm {irr}),
\end{equation} 
where $f$ is replaced by $f_{\rm Bol}$ to be more specific. Equation (7) could be adopted for $V$ apparent visual magnitudes as
\begin{equation}
V = -2.5\times\log (f_{\rm V}) + c_{\rm V}(\rm {irr}),
\end{equation} 
where $c_{\rm V} ({\rm irr})$ is the well-defined zero-point constant for $V$ magnitudes, and $f_{\rm V}$ is the $V$ filtered star flux reaching the telescope (if no extinction). Since bolometric flux ($f_{\rm Bol}$) from a star is bigger than its $V$ flux, $c_{\rm Bol} ({\rm irr})> c_{\rm V} ({\rm irr})$. Even if $c_{\rm Bol} ({\rm irr})$ is a negative number (numerical value of $f_{\rm Bol}$ is less than one), $c_{\rm V} ({\rm irr})$  must also be a negative number but a larger absolute value, thus $c_{\rm Bol} ({\rm irr})$ minus $c_{\rm V} ({\rm irr})$ is always positive. Subtracting equation (8) from (7), the following equation is written. 

\begin{equation}
BC_{\rm V}= m_{\rm Bol}-V = 2.5\times\log (f_{\rm V}/f_{\rm Bol}) + c_{\rm Bol} - c_{\rm V}.
\end{equation} 
Since both $c_{\rm Bol}$ and $c_{\rm V}$ are well-defined constants, equation (9) requires as well that the zero point of the ${\rm BC}_{\rm V}$ scale is also not arbitrary. Even though the zero-point constants are not the same as in equation (4), where they were defined to give luminosities, here, they are defined to give irradiance fluxes reaching the telescope, and the zero-point constant $C_2$, which is defined before as $C_{\rm Bol}-C_{\rm V}$, has the same value as $C_2=c_{\rm Bol}-c_{\rm V}$ in equation (9). That is because the left hand (${\rm BC}_{\rm V}$) side of both equations are the same and  

\begin{equation}
\frac{f_{\rm V}}{f_{\rm Bol}}=\frac{L_{\rm V}}{L}.
\end{equation} 

Despite the fact that the zero-point constants in (2), (3), and (4) are different than the zero-point constants in equations (7), (8), and (9), one ends up with the same numerical value for $C_2$. Taking it as $C_2= c_{\rm Bol}-c_{\rm V}$ from the equation (9), and replacing $f_{\rm V}/f_{\rm Bol}$ by $L_{\rm V}/L$ according to (10), the following equation could be written:
\begin{equation}
L_{\rm V}= L\times 10^{({\rm BC}_{\rm V}-C_2)/2.5}.
\end{equation} 

This is the very same equation used above (equation 5) as an argument for superseding the paradigms of ``bolometric corrections must always be negative'' and ``bolometric magnitude of a star ought to be brighter than its visual magnitude''. It is clear that fixing $c_{\rm Bol}=-18.997~351~...$ by IAU 2015 GAR B2 not only breaks the paradigm of arbitrary zero-point for apparent bolometric magnitudes but also breaks the paradigm of arbitrary zero-point for the $BC_{\rm V}$ scale. It also breaks the other two paradigms. Since the value of $c_{\rm Bol}$ is derived from the value of $C_{\rm Bol}$, as explained in IAU 2015 GAR B2, fixing $C_{\rm Bol}$ is sufficient to break all three paradigms in a single step. That is so whether the BC value of the star is coming from its apparent (equation 9) or its absolute (equation 4) magnitude. 

Note, that the $V$ filter here represents any filter in any photometric system. The equations written for the $V$ filter are also valid for other filters with proper BC and $C_2$ in any photometric system.

\section{Discussions}
\subsection{Vega System of Magnitudes and BC$_{\zeta}$}

One counter argument used  for advocating the arbitrariness of the BC scale is that IAU 2015 GAR B2 did not set the bolometric *correction* scale. It defined only the bolometric magnitude scale set to SI irradiance ($m_{\rm Bol}$, $f_{\rm Bol}$) and luminosity ($M_{\rm Bol}$, $L_{\rm Bol}$) values, whereas the photometric magnitudes are independently defined by standard stars (for Vega magnitudes, or fluxes for ST or AB scale). The IAU could easily set the former, but not the later like the Johnson $V$ system. Since IAU did not set the BC scale, nothing new could be claimed about it, especially about the arbitrariness attributed to the BC scale which has been used since about eight decades.

Such an argument advocating the BC system used before 2015 since about eight decades, is invalid anymore because: equation (9) could be generalized for any photometric system: e.g. Johnson system of apparent magnitudes ($U$, $B$, $V$, $R$, $I$) as 

\begin{eqnarray}
m_{\rm Bol}= \zeta + {\rm BC}_{\zeta} = U+{\rm BC}_{\rm U}= B+{\rm BC}_{\rm B}\\ \nonumber
=V+{\rm BC}_{\rm V} = R+{\rm BC}_{\rm R}= I+{\rm BC}_{\rm I}
\end{eqnarray} 
and equation (4) could be generalized to include the Johnson system of absolute magnitudes ($M_{\rm U}$, $M_{\rm B}$, $M_{\rm V}$, $M_{\rm R}$, $M_{\rm I}$) as 
\begin{eqnarray}
M_{\rm Bol}= M_{\zeta} + {\rm BC}_{\zeta} = M_{\rm U} + {\rm BC}_{\rm U}= M_{\rm B}+{\rm BC}_{\rm B}\\ \nonumber
=M_{\rm V}+{\rm BC}_{\rm V}= M_{\rm R}+{\rm BC}_{\rm R}= M_{\rm I}+{\rm BC}_{\rm I}
\end{eqnarray} 
where subscripts indicate a filter in a photometric system. Apparently, whether it was intentional or not, the resolution conducted by IAU 2015 GAR B2 was the most easiest and logical one. This is because it is illogical to set innumerable zero points for the BC of each band in various photometric systems while there exists an easy way to resolve the problem in a single step. It is obvious in the equations above that fixing the zero point of bolometric magnitudes by equation (6) (or by equation 1) is sufficient to fix the zero points of $\rm BC_{\zeta}$ at once because the Vega system of magnitudes has already well defined zero points as expressed by equation (8). The arbitrariness of the BC scale was removed automatically together with the arbitrariness of the bolometric magnitude system at the moment IAU 2015 GAR B2 was issued. No extra setting was required.

IAU 2015 GAR B2 is revolutionary in part because it opens new insight into fundamental astrophysics, resolving chronic problems of lasting legacy while providing new opportunities towards most accurate stellar $L$. Accuracy may even surpass the direct method \citep{Eker21b}, which can provide typical accuracy of 8.2-12.2\% based on accurate stellar $R$ and $T_{\rm eff}$.

Indicated by equal signs, It is obvious in equations (12, 13) that measurements at any filter are independent from the measurements at other filters. That is, it is possible to obtain independent BC-$T_{\rm eff}$ relations at various filters using the most accurate $R$ and $T_{\rm eff}$ of Double-lined Detached Eclipsing Binaries (DDEB) by the methods of \citet{Flower96} and \citet{Eker20}. Independent relations at {\it Gaia} filters ($G$, $G_{\rm BP}$, $G_{\rm RP}$) and Johnson $B$ and $V$ were obtained by \citet{Bakis22}. Coefficients and basic statistics of five independent BC-$T_{\rm eff}$ relations (fourth degree polynomials) are given in Table 1, where the columns are order, band, coefficients and associated errors to define  
\begin{equation}
{\rm BC}=a+bX+cX^2+dX^3+eX^4
\end{equation}
where $X=\log T_{\rm eff}$, number of DDEB stars ($N$), standard deviation (SD) and correlation coefficient ($R^2$).  

\begin{table*}[!ht]
    \centering
\caption{Coefficients of multiband $BC$-$T_{\rm eff}$ relations taken from \citet{Bakis22}.}
    \begin{tabular}{cccccccccc}
    \hline
Order & Band  & $a$ & $b$  & $c$ & $d$  & $e$  & $N$ & SD & $R^2$ \\
\hline
1 & {\it Gaia} $G$  & -1407.14   & 1305.08    & -453.605 & 70.2338 & -4.1047 & 402 & 0.1108 & 0.9793 \\ 
  &           & $\pm$256.7 & $\pm$258.9 &$\pm$97.67 &$\pm$16.34 &$\pm$1.023 &   &   &   \\ 
\hline
2 & {\it Gaia} $G_{\rm BP}$ & -3421.55 &    3248.19 &   -1156.82 & 183.372 & -10.9305 & 402 & 0.1266 & 0.9738 \\

  &   &$\pm$293.6 &$\pm$296.1 &$\pm$111.7 &$\pm$18.68 &$\pm$1.169 &   &   &   \\ 
\hline
3 & {\it Gaia} $G_{\rm RP}$ &    -1415.67 &    1342.38 &   -475.827 & 74.9702 & -4.44923 & 402 & 0.1092 & 0.9884 \\
  &   &$\pm$253.3 &$\pm$255.4 &$\pm$96.34 &$\pm$16.12 &$\pm$1.009 &   &   &   \\ 
\hline
4 & Johnson $B$ &    -1272.43 &    1075.85 &   -337.831 & 46.8074 & -2.42862 & 342 & 0.1363 & 0.9616 \\ 
          &   &$\pm$394.2 &$\pm$394.4 &$\pm$147.7 &$\pm$24.53 &$\pm$1.152 &   &   &   \\ 
\hline
5 & Johnson $V$ &    -3767.98 &    3595.86 &   -1286.59 & 204.764 & -12.2469 & 386 & 0.1201 & 0.9788 \\ 
 &   &$\pm$288.8 &$\pm$290.9 &$\pm$109.6 &$\pm$18.32 &$\pm$1.146 &   &   &   \\ 
 \hline
\end{tabular}
\end{table*}

Fig. 1 displays the curves of BC-$T_{\rm eff}$ relations in Table 1 as functions of $\log T_{\rm eff}$. Note all five curves cross over one another, as displayed, at $T_{\rm eff}\cong 10000$ K. This is not a coincidence. It is a natural result of using Vega system of magnitudes. 

Using $\alpha$ Lyr (Vega) as the primary calibrating star, the VEGA system is the most well known and deliberated for heterochromatic measurements. Although the zero points are often determined observationally from a network of standard stars, it is formally just a single object. A hypothetical star of
the spectral type A0V with magnitude $V=0.0$ mag on the Johnson system is given in Table 16.6 \citep{Cox00}, where {\it UBVRI} bands and monochromatic fluxes at effective wavelengths are presented. Note that Vega is used as a calibrating star but its apparent magnitude is not exactly zero. $V=0.03$ mag has been measured by \citet{Johnson66} and \citet{Bessell98}. The standard Johnson value of $V=0.03$ mag is cited by \citet{Bohlin14}. The same value ($V=0.03$ mag) was adopted by \citet{Cox00}, \citet{Girardi02}, \citet{Bessell12} and \citet{Casagrande14}. For the other bands, Vega is found to be just slightly positive (0.02 mag) at most bands \citep{Rieke08}. For the heterochromatic bands of various photometric systems, equation (8) could be written as:

\begin{equation}
\zeta = -2.5\times\log (f_{\rm \zeta}) + c_{\rm \zeta}(\rm {irr}),
\end{equation} 
where the zero point constant $c_{\rm \zeta}(\rm {irr})$ needs to be derived for each bandpass $\zeta$ using a star of known absolute flux, usually Vega \citep{Casagrande14}, or Sirius and Vega \citep{Bohlin14}. Although $c_{\rm \zeta}(\rm {irr})$ are usually not given for the photometric systems in the literature, where only monochromatic fluxes making apparent magnitudes zero at effective wavelengths of the filters are listed, all of the zero point constants associated with Vega system of magnitudes at various bands of different photometric systems are well-defined quantities \citep{Bessell98, Cox00, Girardi02, Casagrande14}. 

Figure 1 confirms usage of Vega system of magnitudes through equations (12) and (13) because ${\rm BC} - T_{eff}$ curves cut each other at a single point ($T_{\rm eff}\cong 10000$ K) as the color curves would do the same since the colors of the hypothetical star A0V is taken to be $U-B=B-V=V-R=V-I=0$ mag. Curves intercepting at single point is naturally expected because the difference of ${\rm BC}_\zeta$ values at two different bands is equal to the negative value of the colors defined in the photometric system according to the basic definition of BCs. All of these could be counted as undeniable evidences indicating the zero point constants of ${\rm BC}_\zeta$ are not arbitrary but well defined constants as the zero point constants of Vega system of magnitudes.

\begin{figure*}
\centering
\includegraphics[scale=0.40, angle=0]{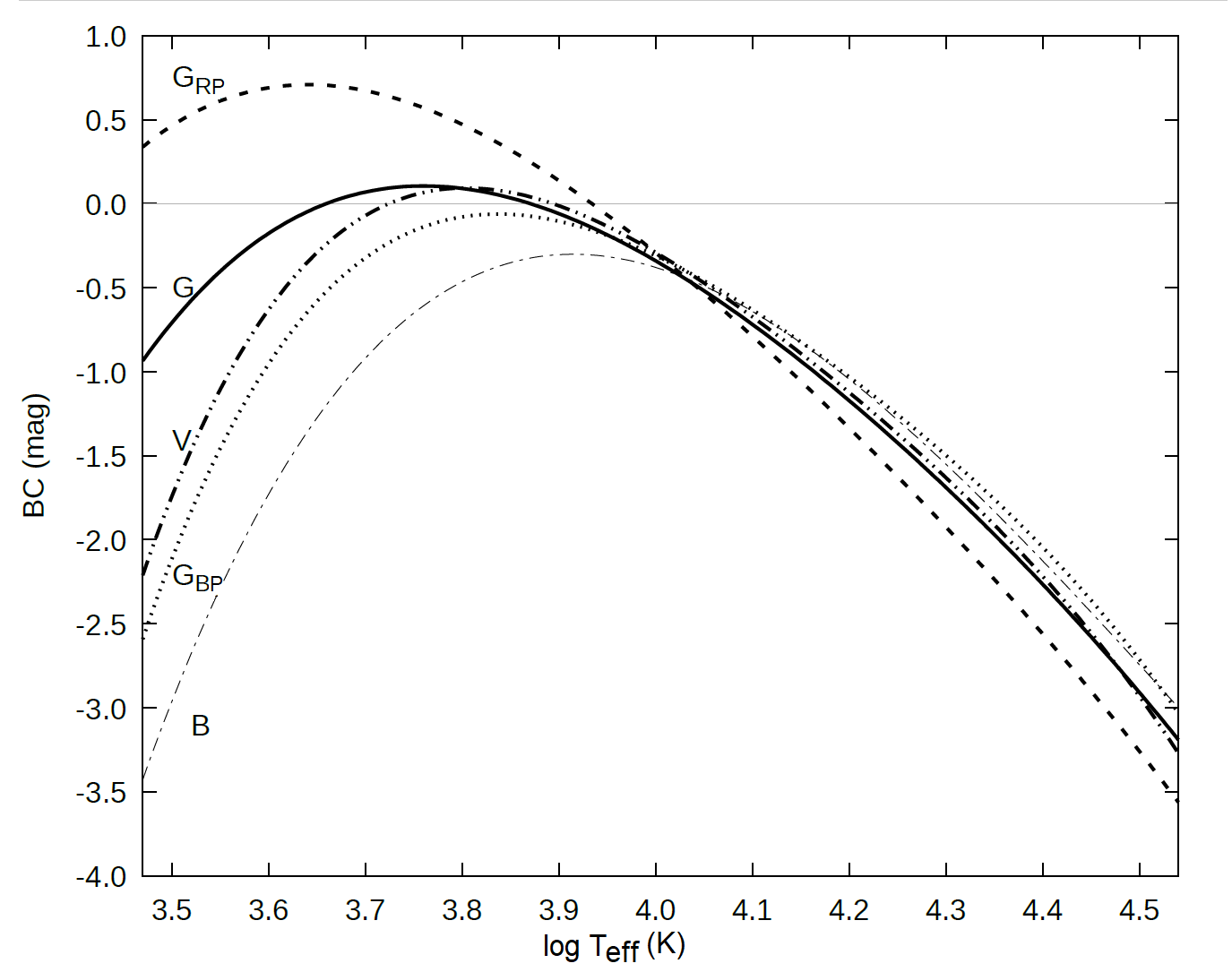}
\caption{Independent BC-$T_{\rm eff}$ relation at {\it Gaia} $G$, $G_{\rm BP}$, $G_{\rm RP}$ and Johnson $B$ and $V$ \citep{Bakis22}.} 
\end{figure*}

Each curve in Fig. 1 could be used to estimate a BC of a star if its effective temperature is known. Then this BC could be used to calculate its bolometric magnitude as $M_{\rm Bol} = M_{\zeta}+ {\rm BC}_{\zeta}$ if its distance and $E(B-V)$ color excess are known. At last, its standard luminosity ($L$) could be calculated from $M_{\rm Bol}$ using  equation (2). The accuracy of $M_{\rm Bol}$, and accuracy of $L$ depend on the accuracies of $M_{\zeta}$ and ${\rm BC}_{\zeta}$ if standard BC values are used, otherwise zero-point errors ($\sim 10\%$) caused by arbitrary definitions \citep{Torres10, Eker21a, Eker21b} must also be added. SD column in Table 1 indicates typical accuracy of a BC. That is if $M_{\zeta}$ is errorless it is possible to obtain $M_{\rm Bol}$ with accuracy $\pm$SD, which means $\Delta L/L\sim 0.921\times$SD \citep{Eker21b}. Thus according to Table 1 the most accurate $L(\sim 10\%)$ could be obtained using BC from BC-$T_{\rm eff}$ relation at the band {\it Gaia} $G_{\rm RP}$, obviously not better than the direct method.   

Most important is that one may increase the accuracy of predicted $L$ if many independent standard BC-$T_{\rm eff}$ relations are used in. For example, the primary of HP Aur has $M=0.9543\pm0.0041 M_{\odot}$ $R=1.0278\pm0.0042 R_{\odot}$ and $T_{\rm eff}=5810\pm120$ K \citep{Lacy14}. The Stefan-Boltzmann law gives its $L=4.149\pm0.344\times 10^{26}$ W, with an accuracy of 8.302\%. Using interstellar extinctions $A_{\rm V}=0.335$, $A_{\rm G}=0.298$, $A_{\rm BP}=0.366$, $A_{\rm RP}=0.207$ mags and its {\it Gaia} EDR3 trigonometric parallax $5.2432\pm 0.0306$ mas \citep{Gaia21}, \citet{Bakis22} calculated its absolute magnitudes $M_{\rm V}=4.753\pm 0.033$, $M_{\rm G}=4.583\pm 0.033$, $M_{\rm BP}=4.860\pm 0.042$, and $M_{\rm RP}=4.152\pm 0.024$ mags. Using the BC values from Table 1, its bolometric absolute magnitudes were found 4.824, 4.688, 4.730, and 4.713 mags, which are then combined to a single value as $4.739\pm 0.030$ mag. At last its luminosity of $L=3.831\pm 0.096\times 10^{26}$ W is found with an uncertainty of 2.5\%. That is about four times better than the direct method. Hitherto, no method available could provide a stellar luminosity more accurate than the direct method. This is the first, and became possible only after IAU 2015 GAR B2. Accurate stellar luminosities are not only needed to test stellar structure and evolution theories but also required for improving Galactic and extra-galactic studies, dark matter search and even perhaps to be used to resolve the Hubble tension at last because stars are one of the most important primary 
building blocks to understand the universe.

\subsection{Attention Required On Zero-Point Constants}

The apparent magnitude of a star disregards distance information. Thus, equation (7) is valid only for the apparent magnitudes of stars. The very same equation could be adopted for absolute bolometric magnitudes as 

\begin{equation}
M_{\rm Bol}= -2.5\times\log (F_{\rm Bol}) + c_{\rm Bol}(\rm {irr}),
\end{equation} 
on the condition that the star is assumed to be at a fixed distance of 10 pc. It is only under this condition that an irradiance or heat flux density of $F_{\rm Bol}=f_0=2.518021002~...\times 10^{-8}$ W m$^{-2}$ at the focal point of the telescope, makes it possible to write $M_{\rm Bol}=0$ mag. It is obvious that this equation and equation (2) appear different despite the left-hand side of both equations being the same. Same $M_{\rm Bol}$ before the equal sign implies that 
\begin{equation}
-2.5\times\log (L) + C_{\rm Bol}= -2.5\times\log (F_{\rm Bol}) + c_{\rm Bol}(\rm {irr}).
\end{equation} 

The algebraic sum of the two quantities on the right-hand side of both equations must be the same. Thus, the zero-point constant associated with luminosities is different than the zero-point constant associated with irradiances. 

Subtracting equation (16) from equation (7), which has the same zero-point constant, they must cancel in the subtraction, thus

\begin{equation}
m_{\rm Bol}-M_{\rm Bol}=2.5\times\log (F_{\rm Bol}/f_{\rm Bol}).
\end{equation} 
Because $F_{\rm Bol}$ and $f_{\rm Bol}$ correspond to bolometric fluxes of the same star if it is at a distance of 10 pc and d pc away (assuming no extinctions) and the flux of a star is inversely proportional to the square of its distance, then $(F_{\rm Bol}/f_{\rm Bol})=(d/10)^2$. Inserting this into (18),

\begin{equation}
m_{\rm Bol}-M_{\rm Bol}=5\times\log d-5,
\end{equation} 
the distance modulus of the star is obtained. So, one has to be careful using absolute magnitudes. Proper usage of the zero-points is necessary. If $M_{\rm Bol}$ is associated with $L$, then use $C_{\rm Bol}$ as in equation (2). Else, use $c_{\rm Bol}{\rm (irr)}$ as in equation (16).

\section{Conclusions}

The results obtained in this study are listed below:
\begin{itemize}

\item Although its primary aim was to set the zero point constant of bolometric magnitudes, IAU 2015 GAR B2 was shown to be  a revolutionary document to imply the zero point constants of ${\rm BC_{\rm \zeta}}$, where $\zeta$ indicates various photometric bands, are not arbitrary but well defined constants.

\item IAU 2015 GAR B2 is a revolutionary document not only because it solves nearly a century old problems caused by the arbitrariness attributed to the BC scale, but also to indicate ``bolometric corrections must always be negative'' and ``bolometric magnitude of a star ought to be brighter than its $V$ magnitude'' are not right anymore.

\item The falsehood of the three paradigms (BC scale is arbitrary, BC values must always be negative, bolometric magnitude of a star ought to be brighter than its $V$ magnitude) are not only proven mathematically but also confirmed from observational point of view by multiband ${\rm BC}-T_{\rm eff}$ relations calibrated by \citet{Bakis22} who used published observational data of DDEB stars. 

\item Increasing the number of photometric bands with standard ${\rm BC}-T_{\rm eff}$ relations, which is to be used in calculating standard stellar luminosity from $M_{\rm Bol}= M_{\zeta}+ {\rm BC}_{\zeta}$ and $M_{\rm Bol}=-2.5\times \log L + C_{\rm Bol}$, increases the accuracy of the standard luminosity of a star. Using this method in predicting stellar luminosities is also a revolution because this method is shown to be predicting stellar luminosities much more accurate than the classical direct method using stellar radii and effective temperatures through the Stefan Boltzmann law. Achieving such accuracy became possible only after discarding the arbitrary zero point errors caused by the arbitrariness attributed to the BC scale; this was possible, however, only after understanding the potential of IAU 2015 GAR B2.

\item Such an accuracy in calculated stellar luminosities is not only needed by stellar structure and evolution theories to be refined but also needed by Galactic and extragalactic studies including observational and theoretical cosmology, dark matter search etc to be improved further since stars are primary building blocks of the universe. 

\end{itemize}

\section{Acknowledgements}
This work has been supported in part by the Scientific and Technological Research Council (T\"UB\.ITAK) under grant number 114R072.
This research has made use of NASA's Astrophysics Data System Bibliographic Services. This research made use of VizieR and Simbad databases at CDS, Strasbourg, France. We made use of data from the European Space Agency (ESA) mission \emph{Gaia}\footnote{https://www.cosmos.esa.int/gaia}, processed by the \emph{Gaia} Data Processing and Analysis Consortium (DPAC)\footnote{https://www.cosmos.esa.int/web/gaia/dpac/consortium}. Funding for DPAC has been provided by national institutions, in particular the institutions participating in the \emph{Gaia} Multilateral Agreement.


\end{document}